\documentclass[10pt]{article}
\usepackage{graphicx}
\usepackage{amsmath}
\usepackage{amssymb}
\usepackage{caption2}
\begin{document}
\bibliographystyle{prsty}
\begin{center}
{\large {\bf \sc{  Analysis of the decay constants of the heavy pseudoscalar mesons with  QCD sum rules }}} \\[2mm]
Zhi-Gang Wang \footnote{E-mail,zgwang@aliyun.com.  }     \\
 Department of Physics, North China Electric Power University,
Baoding 071003, P. R. China
\end{center}

\begin{abstract}
In this article, we recalculate the contributions of all   vacuum condensates up to dimension-6, in particular the one-loop corrections  to the quark condensates $\alpha_s\langle \bar{q}q\rangle$ and partial one-loop corrections to the four-quark condensates  $\alpha_s^2\langle \bar{q}q \rangle^2$, in the operator product expansion.  Then we study the masses and decay constants of the heavy pseudoscalar mesons $D$, $D_s$, $B$ and $B_s$ using the QCD sum rules with  two choices:
{\bf I } we choose the $\overline{MS}$ masses by setting $m=m(\mu)$ and take perturbative corrections up to the order $\mathcal{O}(\alpha_s)$;
{\bf II} we choose the pole masses $m$, take perturbative corrections up to the order $\mathcal{O}(\alpha_s^2)$ and set the energy-scale to be the heavy quark pole mass $\mu=m_Q$. In the case of {\bf I}, the predictions  $f_D=(208\pm11)\,\rm{MeV}$ and $f_B=(189\pm15)\,\rm{MeV}$ are consistent with the experimental data within uncertainties, while the prediction $f_{D_s}=(241\pm12)\,\rm{MeV}$ is  below the lower bound of the experimental data $f_{D_s}=(260.0\pm5.4)\,\rm{MeV}$.
In the case of {\bf II}, the predictions $f_D=(211\pm14)\,\rm{MeV}$, $f_B=(190\pm17)\,\rm{MeV}$, $f_{D_s}=(258\pm13)\,\rm{MeV}$ and $f_{D_s}/f_D=1.22\pm0.08$ are all in excellent agreements  with the experimental data within uncertainties.
\end{abstract}



\section{Introduction}
The charged pseudoscalar mesons $\pi^+$, $K^+$, $D_s^+$ and $B^+$ mesons  can decay to a charged lepton pair $\ell^+ {\nu}_\ell$ through  a virtual
$W^+$ boson. To the lowest order, the decay width is
\begin{eqnarray}
\Gamma(P\to \ell\nu) &=& {{G_F^2}\over 8\pi}f_{P}^2  m_{\ell}^2m_{P}\left(1-{m_{\ell}^2\over m_{P}^2}\right)^2 \left|V_{q_1q_2}\right|^2\, ,
\end{eqnarray}
where the $m_{P}$ and $f_P$ are the mass and decay constant of  the pseudoscalar meson, respectively,  the $m_{\ell}$ is the $\ell$
mass, the $V_{q_1 q_2}$ is the Cabibbo-Kobayashi-Maskawa  matrix element between the constituent quarks $q_1\bar{q}_2$, and the $G_F$ is the Fermi coupling constant.
The CLEO collaboration obtains the values  $f_D=(202 \pm 41 \pm 17)\, \rm{MeV}$ \cite{CLEO-2004-fD},  $(222.6 \pm 16.7 {}^{+2.8}_{-3.4})\,\rm{ MeV}$ \cite{CLEO-2005-fD}, $(205.8 \pm 8.5 \pm 2.5)\,\rm{ MeV}$ \cite{CLEO-2008-fD} from the decay $D^+ \to \mu^+ \nu_\mu$;  $f_{D_s}=(259.5 \pm 6.6 \pm 3.1)\,\rm{ MeV}$,
$f_{D_s}/f_{D}=1.26 \pm 0.06 \pm 0.02$ \cite{CLEO-2009-fDs} from the decay $D_s^+ \to \mu^+ \nu_\mu$; $ f_{D_s} = (252.5 \pm 11.1 \pm 5.2)\,\rm{MeV}$ \cite{CLEO-2009-fDs-2}, $ (259.7 \pm 7.8 \pm 3.4)\,\rm{MeV}$ \cite{CLEO-2009-fDs-3} from the decay $D_s^+ \to \tau^+ \nu_\tau$. The  BaBar  collaboration obtains  the value $f_{D_s} =
(258.6 \pm 6.4 \pm 7.5)\,\rm{ MeV}$ \cite{BaBar-2010-fDs} from the decays $D_s^- \to \ell^- \bar{\nu}_\ell$. The Belle collaboration obtains the value $f_{D_s} = (275 \pm 16  \pm 12\, \rm{MeV} $ \cite{Belle-2008-fDs} from the decay $D_s^+\to\mu^+\nu_\mu $. Now the average values listed in the  Review of Particle Physics
 are $f_D=(206.7 \pm 8.9)\,\rm{MeV}$, $f_{D_s}=(260.0 \pm 5.4)\,\rm{MeV}$ and $f_{D_s}/f_D=1.26 \pm 0.06$ \cite{PDG}.

There have been many theoretical works on the decay constants of the heavy pseudoscalar mesons, such as the QCD sum rules (QCDSR) \cite{Aliev-pert,QCDSRfD1,QCDSR-3loop,SRfDEFT1,QCDSRfD2,fD-Narison-2008,QCDSRfD3,QCDSRfD4,QCDSRfD5}, the lattice QCD (LQCD) \cite{LattfD,LattfD1,LattfD2,LattfD3}, the Bethe-Salpeter equation (BSE) \cite{BSEfD1,BSEfD2}, the relativistic potential model (RPM) \cite{RPMfD1,RPMfD2,RPMfD3},  the field-correlator method (FCM) \cite{FCMfD1}, the light-front quark model (LFQM) \cite{LFQMfD1,LFQMfD2}, the chiral extrapolation \cite{Guo-fD}, the extended  chiral-quark model \cite{ChQM-fD}, etc. There are discrepancies between the theoretical values (from QCDSR and LQCD) and experimental data, which maybe signal some new physics beyond the standard model \cite{fD-Narison-2008}. In the QCD sum rules for the heavy pseudoscalar mesons, the Wilson coefficients of the vacuum condensates  at the operator product expansion side from different references are different from each other in one way or the other, as  different  authors take different approximations in their calculations  \cite{QCDSRfD1,QCDSRfD2,QCDSRfD5,QCDSR-mass}.

In this article, we recalculate the contributions of all  vacuum condensates up to dimension-6, in particular the one-loop corrections  to the quark condensates $\alpha_s\langle \bar{q}q\rangle$ and partial one-loop corrections to the four-quark condensates  $\alpha_s^2\langle \bar{q}q \rangle^2$, in the operator product expansion, take into account all  terms neglected in previous works,
then study the masses and decay constants of the heavy pseudoscalar mesons $D$, $D_s$, $B$ and $B_s$ with the QCD sum rules. The QCD sum rules is a powerful theoretical tool  in studying   the ground state hadrons \cite{SVZ79,Reinders85}.
 The vacuum condensates play an important role in determining the Borel windows, although they maybe play a less important role in the Borel windows. Different  Borel windows lead to different ground state  masses, therefore different decay constants.

The article is arranged as follows:  we derive the QCD sum rules for
the masses and decay constants of the heavy pseudoscalar  mesons  in Sect.2;
in Sect.3, we present the numerical results and discussions; and Sect.4 is reserved for our
conclusions.

\section{QCD sum rules for  the heavy pseudoscalar  mesons }
In the following, we write down  the two-point correlation functions
$\Pi(p)$  in the QCD sum rules,
\begin{eqnarray}
\Pi(p)&=&i\int d^4x e^{ip \cdot x} \langle
0|T\left\{J_{5}(x)J_{5}^{\dagger}(0)\right\}|0\rangle \, , \\
J_{5}(x)&=&\bar{Q}(x)i\gamma_5 q(x)  \, ,
\end{eqnarray}
where the pseudoscalar currents $J_{5}(x)$  interpolate the heavy pseudoscalar mesons, $Q=c,b$ and $q=u,d,s$.
We can insert  a complete set of intermediate hadronic states with
the same quantum numbers as the current operators $J_{5}(x)$ into the
correlation functions $\Pi(p)$  to obtain the hadronic representation
\cite{SVZ79,Reinders85}. After isolating the ground state
contributions from the heavy pseudoscalar mesons, we get the following result,
\begin{eqnarray}
\Pi(p)&=&\frac{f_{P}^2m_{P}^4}{(m_Q+m_q)^2(m_{P}^2-p^2)} +\cdots\,  ,
\end{eqnarray}
where the  decay constants $f_{P}$ are defined by
\begin{eqnarray}
\langle 0|J_{5}(0)|P(p)\rangle&=&\frac{f_{P}m^2_{P}}{m_Q+m_q}  \, .
\end{eqnarray}

Now, we briefly outline  the operator product
expansion for the correlation functions $\Pi(p)$  in perturbative
QCD, and use the charm-strange (or bottom-strange) mesons to illustrate the  procedure.  We contract the quark fields in the correlation functions
$\Pi(p)$ with Wick theorem firstly,
\begin{eqnarray}
\Pi(p)&=&i\int d^4x e^{ip \cdot x}   Tr\left\{\gamma_{5}S_{ij}(x)\gamma_{5} S^Q_{ji}(-x) \right\}\, ,
\end{eqnarray}
where the $S_{ij}(x)$ and $S^Q_{ij}(x)$ are the full quark propagators, and can be written as
\begin{eqnarray}
S_{ij}(x)&=& \frac{i\delta_{ij}\!\not\!{x}}{ 2\pi^2x^4}
-\frac{\delta_{ij}m_s}{4\pi^2x^2}-\frac{\delta_{ij}}{12}\langle
\bar{s}s\rangle +\frac{i\delta_{ij}\!\not\!{x}m_s
\langle\bar{s}s\rangle}{48}-\frac{\delta_{ij}x^2\langle \bar{s}g_s\sigma Gs\rangle}{192}+\frac{i\delta_{ij}x^2\!\not\!{x} m_s\langle \bar{s}g_s\sigma
 Gs\rangle }{1152}\nonumber\\
&& -\frac{iG^{a}_{\alpha\beta}t^a_{ij}(\!\not\!{x}
\sigma^{\alpha\beta}+\sigma^{\alpha\beta} \!\not\!{x})}{32\pi^2x^2} +\frac{i\delta_{ij}x^2\!\not\!{x}g_s^2\langle \bar{s}\gamma_\mu t^n s\bar{s}\gamma^\mu t^n s\rangle}{3456}   +\cdots \, ,
\end{eqnarray}
\begin{eqnarray}
S^Q_{ij}(x)&=&\frac{i}{(2\pi)^4}\int d^4k e^{-ik \cdot x} \left\{
\frac{\delta_{ij}}{\!\not\!{k}-m_Q}
-\frac{g_sG^n_{\alpha\beta}t^n_{ij}}{4}\frac{\sigma^{\alpha\beta}(\!\not\!{k}+m_Q)+(\!\not\!{k}+m_Q)
\sigma^{\alpha\beta}}{(k^2-m_Q^2)^2}\right.\nonumber\\
&&\left. +\frac{g_s D_\alpha G^n_{\beta\lambda}t^n_{ij}(f^{\lambda\beta\alpha}+f^{\lambda\alpha\beta}) }{3(k^2-m_Q^2)^4}-\frac{g_s^2 (t^at^b)_{ij} G^a_{\alpha\beta}G^b_{\mu\nu}(f^{\alpha\beta\mu\nu}+f^{\alpha\mu\beta\nu}+f^{\alpha\mu\nu\beta}) }{4(k^2-m_Q^2)^5}+\cdots\right\} \, ,\nonumber\\
f^{\lambda\alpha\beta}&=&(\!\not\!{k}+m_Q)\gamma^\lambda(\!\not\!{k}+m_Q)\gamma^\alpha(\!\not\!{k}+m_Q)\gamma^\beta(\!\not\!{k}+m_Q)\, ,\nonumber\\
f^{\alpha\beta\mu\nu}&=&(\!\not\!{k}+m_Q)\gamma^\alpha(\!\not\!{k}+m_Q)\gamma^\beta(\!\not\!{k}+m_Q)\gamma^\mu(\!\not\!{k}+m_Q)\gamma^\nu(\!\not\!{k}+m_Q)\, ,
\end{eqnarray}
and  $t^n=\frac{\lambda^n}{2}$, the $\lambda^n$ is the Gell-Mann matrix, the $i$, $j$ are color indexes, $D_\alpha=\partial_\alpha-ig_sG^n_\alpha t^n$ \cite{Reinders85}; then compute  the integrals both in
the coordinate and momentum spaces;  finally obtain the correlation functions $\Pi(p)$ at the
level of   quark-gluon degrees  of freedom. In Figs.1-4, we express the contributions of the mixed condensates, four-quark condensates, gluon condensates and three-gluon condensates in terms of Feynman diagrams, which are drawn up directly from Eqs.(6-8). In the Feynman diagrams, we use the solid and dashed lines to represent the light and heavy quark propagators, respectively.

The analytical expressions of the perturbative $\mathcal{O}(\alpha_s)$ corrections \cite{Aliev-pert} and semi-analytical
expressions of the perturbative $\mathcal{O}(\alpha_s^2)$ corrections \cite{QCDSR-3loop} to the perturbative term are available now. We take  into account those  analytical and semi-analytical expressions directly \cite{Aliev-pert,QCDSR-3loop}; and recalculate the one-loop corrections to the quark condensates. We insert the following term
\begin{eqnarray}
\frac{1}{2!}\,\, ig_s \int d^4 y \bar{\psi}(y)\gamma^\mu \psi(y)t^aG^a_\mu(y)\,\, ig_s \int d^4 z \bar{\psi}(z)\gamma^\nu \psi(z)t^bG^b_\nu(z) \, ,
\end{eqnarray}
 into the correlation functions $\Pi(p)$ firstly, where the $\psi$ denotes the quark fields, then contract the quark fields with Wick theorem,  and extract the quark condensate $\langle\bar{s}{s}\rangle$ according to Eq.(7) to obtain the perturbative corrections $\alpha_s\langle\bar{s}{s}\rangle$. There are six Feynman diagrams make contributions, see Fig.5.
In summary, we calculate the Feynman diagrams shown explicitly in Figs.1-5 to obtain the contributions of the vacuum condensates in the operator product expansion.

In the following, we will present some necessary technical details in calculations. In this article, we take the light quark mass $m_q$ (or $m_s$) as a small quantity and expand it perturbatively. In Fig.5, there exist divergences, the quark condensate in the full propagators should be replaced as
 \begin{eqnarray}
 \frac{\langle \bar{s}s\rangle}{12} \to \frac{\langle \bar{s}s\rangle}{3D}= \frac{\langle \bar{s}s\rangle}{12}\left(1+\frac{1}{2}\epsilon \right) \, .
  \end{eqnarray}
In this article, we carry out the integrals in the dimension $D=4-2\epsilon$ to regularize the divergences, then use the vacuum condensates to absorb the infrared divergences  and choose the on-shell scheme to renormalize the ultraviolet divergences. We can also choose the $\overline{MS}$ scheme to renormalize the ultraviolet divergences, the two schemes are equivalent except that  different masses (pole masses or $\overline{MS}$ masses) are taken.

   In calculations, we observe that the mixed condensates (see Fig.1) are depressed by additional powers of $1/T^2$ compared to the quark condensates. The perturbative $\mathcal{O}(\alpha_s)$ corrections to the mixed condensates are doubly depressed by the factor $\alpha_s/T^2$ and play a less important role,  they are neglected in this article.  In the massless limit, the second Feynman diagram in Fig.2 does not contribute to the gluon condensate $\langle \frac{\alpha_sGG}{\pi}\rangle$, the QCD spectral density is $\frac{1}{8}\langle \frac{\alpha_sGG}{\pi}\rangle$  instead of $\frac{1}{12}\langle \frac{\alpha_sGG}{\pi}\rangle$. In few articles, the coefficient of the gluon condensate is taken as $\frac{1}{8}$ regardless of the heavy quark masses. In calculating the fifth Feynman diagram in Fig.2, we use the light quark propagator $S(k)$ in the momentum space,
 \begin{eqnarray}
S(k)&=&-\frac{\langle g_s^3GGG\rangle}{24}\frac{D^3-9D^2+20D-12}{D(D-1)(D-2)}\frac{i\!\not\!{k}}{k^8}=\frac{\langle g_s^3GGG\rangle}{48}\left(1+\frac{3}{2}\epsilon\right) \frac{i\!\not\!{k}}{k^8} \, .
\end{eqnarray}
  The expression presented in Ref.\cite{Reinders85} is correct only in four-dimension, we add the factor $\frac{3}{2}\epsilon$ to obtain the propagator in
 $D$-dimension as there are divergences; in few articles, the $\epsilon$ is discarded.  In calculating the Feynman diagrams in Figs.3-4,  we use the equation of motion, $D^{\nu}G_{\mu\nu}^a=\sum_{q=u,d,s}g_s\bar{q}\gamma_{\mu}t^a q $,   and take the approximation  $\langle\bar{s}s\rangle=\langle\bar{q}q\rangle$, furthermore,  we take assumptions of the vacuum saturation and factorization \cite{SVZ79}, and use the following formula,
 \begin{eqnarray}
 \langle \bar{q}\gamma_\mu t^n q\bar{q}\gamma^\mu t^n q\rangle &=&-\frac{16}{9D}\langle \bar{q} q\rangle^2 =-\frac{4}{9}\langle \bar{q} q\rangle^2\left(1+\frac{1}{2}\epsilon\right)\, .
 \end{eqnarray}
 The factor $\frac{1}{2}\epsilon$ cannot be neglected  when companied with divergences  in the loop integral; in few articles, the $\epsilon$ is discarded.  In Fig.4, we present the Feynman Diagram cannot be written as perturbative $\mathcal{O}(\alpha_s)$ corrections to the four-quark condensates $\alpha_s\langle\bar{s}s\rangle^2$ (shown in Fig.3). The four-quark condensates $\alpha_s\langle\bar{s}s\rangle^2$ play a less important role, the perturbative $\mathcal{O}(\alpha_s)$ corrections $\alpha_s^2\langle\bar{s}s\rangle^2$ can be safely neglected, although they appear in one-loop order.

\begin{figure}
 \centering
 \includegraphics[totalheight=3cm,width=14cm]{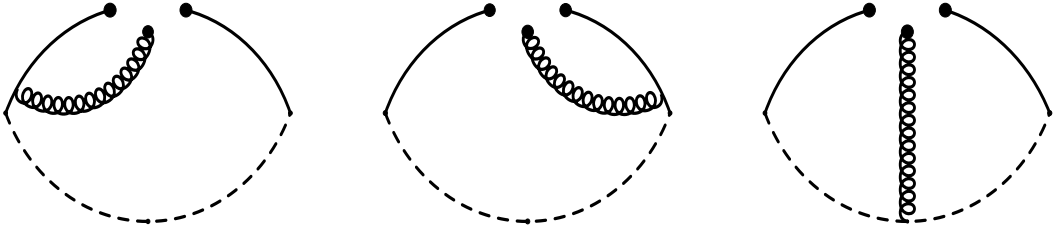}
    \caption{The diagrams contribute to the mixed condensate $\langle\bar{s}g_s \sigma G s\rangle$. }
\end{figure}

\begin{figure}
 \centering
 \includegraphics[totalheight=6cm,width=14cm]{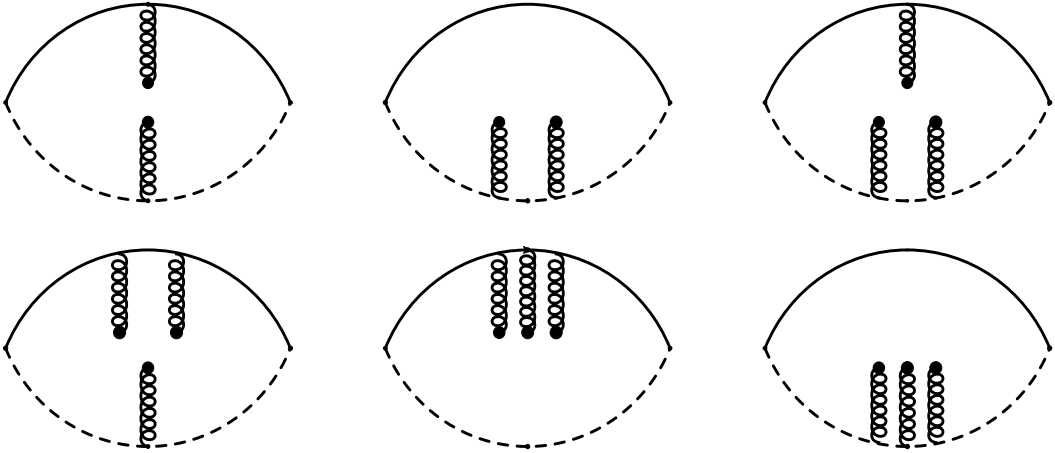}
    \caption{The diagrams contribute to the gluon condensate $\langle \frac{\alpha_sGG}{\pi}\rangle$ and three-gluon condensate $\langle g_s^3 GGG\rangle$. }
\end{figure}

\begin{figure}
 \centering
 \includegraphics[totalheight=3cm,width=14cm]{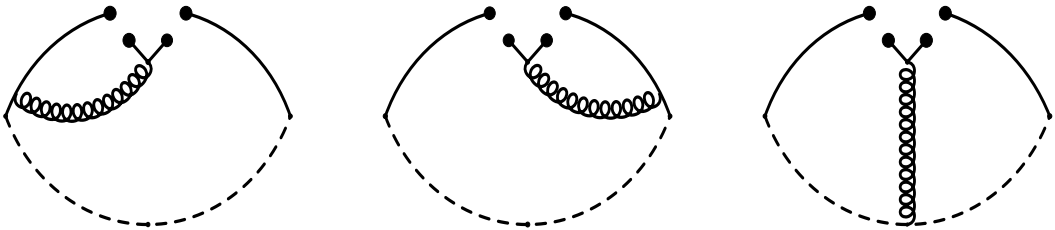}
    \caption{The diagrams contribute to the four-quark condensate $\langle\bar{s} s\rangle^2$ of the order $\mathcal{O}(\alpha_s)$. }
\end{figure}

\begin{figure}
 \centering
 \includegraphics[totalheight=3cm,width=5cm]{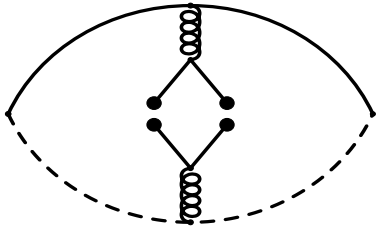}
    \caption{The typical diagram contributes to the four-quark condensate $\langle\bar{s} s\rangle^2$ of the order $\mathcal{O}(\alpha_s^2)$. }
\end{figure}

\begin{figure}
 \centering
 \includegraphics[totalheight=6cm,width=14cm]{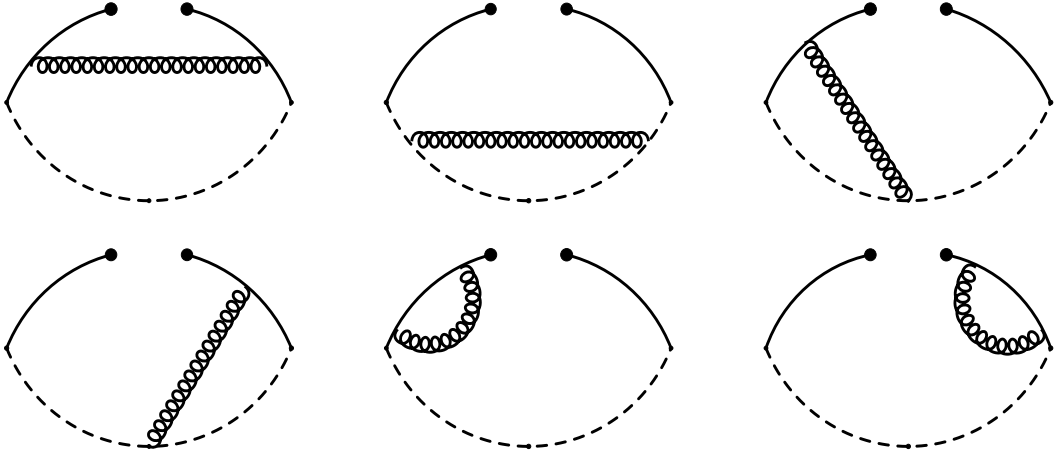}
    \caption{The perturbative $\mathcal{O}(\alpha_s)$ corrections to the quark condensate $\langle\bar{s}s\rangle$. }
\end{figure}

 Once analytical expressions of the QCD spectral densities are obtained,   then we can take the
quark-hadron duality below the continuum thresholds and perform the Borel transforms  with respect to the variable
$P^2=-p^2$ to obtain the QCD sum rules,
\begin{eqnarray}
 \frac{f_{D_s}^2 m_{D_s}^4}{(m_c+m_s)^2}\exp\left(-\frac{m_{D_s}^2}{T^2}\right)&=& \frac{3}{8\pi^2} \int_{m_c^2}^{s_0} ds s\left(1-\frac{m_c^2}{s}\right)^2\left\{1+\frac{2m_s m_c}{s-m_c^2}+\frac{4\alpha_s}{3\pi} R\left(\frac{m_c^2}{s}\right)\right\} \exp\left(-\frac{s}{T^2}\right) \nonumber \\
 &&-m_c\langle\bar{s}s\rangle\left\{1+\frac{\alpha_s}{\pi}\left[ \frac{16}{3}-\frac{5m_c^2}{3T^2}-\left( \frac{4}{3}-\frac{2m_c^2}{3T^2}\right)\log\frac{m_c^2}{\mu^2} -2\exp\left(\frac{m_c^2}{T^2}\right)\right.\right.\nonumber\\
 &&\left.\left. \Gamma\left(0,\frac{m_c^2}{T^2}\right)\right]\right\}\exp\left(-\frac{m_c^2}{T^2}\right)+\frac{m_s\langle\bar{s}s\rangle}{2}\left(1+\frac{m_c^2}{T^2}\right)\exp\left(-\frac{m_c^2}{T^2}\right)\nonumber\\
 && -\left\{\frac{m_c\langle\bar{s}g_s\sigma Gs\rangle}{2T^2}\left(1-\frac{m_c^2}{2T^2}\right)+\frac{m_s m_c^4\langle\bar{s}g_s\sigma Gs\rangle}{12T^6}\right\}\exp\left(-\frac{m_c^2}{T^2}\right) \nonumber\\
 &&+\frac{1}{12}\langle \frac{\alpha_sGG}{\pi}\rangle\exp\left(-\frac{m_c^2}{T^2}\right)-\frac{16\pi\alpha_s\langle\bar{s}s\rangle^2 }{27T^2}\left(1+\frac{m_c^2}{2T^2}
 -\frac{m_c^4}{12T^4}\right)\nonumber\\
 &&\exp\left(-\frac{m_c^2}{T^2}\right) +\frac{\langle g_s^3 GGG\rangle}{\pi^2}\left\{ \frac{65}{4608T^2}+ \frac{77}{4608m_c^2} +\frac{23m_c^2}{1536T^4}+\right.\nonumber\\
 &&\left(\frac{5}{768T^2}+\frac{1}{768m_c^2}-\frac{m_c^2}{512T^4}-\frac{m_c^4}{384T^6} \right)\log \frac{m_c^2\mu^2}{T^4}+\exp\left(\frac{m_c^2}{T^2}\right)\nonumber\\
 &&\left.\Gamma\left(0,\frac{m_c^2}{T^2}\right)\left(\frac{m_c^2}{64T^4}-\frac{m_c^4}{192T^6}-\frac{m_c^6}{768T^8} \right) \right\}\exp\left(-\frac{m_c^2}{T^2}\right)\nonumber\\
 &&+\frac{16\alpha_s^2\langle\bar{q}{q}\rangle^2}{9}\left\{ \frac{1}{6T^2}+\frac{1}{2m_c^2}-\frac{m_c^2}{36T^4}-\left(\frac{1}{6T^2}+\frac{m_c^2}{12T^4} \right) \log \frac{m_c^2\mu^2}{T^4} \right. \nonumber\\
 &&\left.- \frac{1}{3T^2}\exp\left(\frac{m_c^2}{T^2}\right)\Gamma\left(0,\frac{m_c^2}{T^2}\right)\right\}\exp\left(-\frac{m_c^2}{T^2}\right) \, ,
\end{eqnarray}
where
\begin{eqnarray}
R(x)&=&\frac{9}{4}+2{\rm Li}_2(x)+{ \log}x\,{\log}(1-x)-\frac{3}{2}\,{\log}\frac{1-x}{x}-{\log}(1-x)+x\,{\log}\frac{1-x}{x}
-\frac{x}{1-x}{\log}x \, , \nonumber\\
\end{eqnarray}

\begin{eqnarray}
\Gamma(0,x)&=&e^{-x}\int_0^\infty dt \frac{1}{t+x}e^{-t} \, , \nonumber\\
{\rm Li}_2(x)&=&-\int_0^x dt \frac{1}{t} \log(1-t)\, ,
\end{eqnarray}
and the $s_0$ is the continuum threshold parameter.
The perturbative $\mathcal{O}(\alpha_s)$ corrections $R(x)$ are taken from Ref.\cite{Aliev-pert}.
We can also take into account the
semi-analytical perturbative  $\mathcal{O}(\alpha_s^2)$ corrections,
\begin{eqnarray}
\frac{1}{8\pi^2}\left(\frac{\alpha_s}{\pi}\right)^2\int_{m_c^2}^{s_0}ds \left\{ \frac{16}{9}\,{\rm R2sFF}[v]+4\,{\rm R2sFA}[v]+\frac{2n_l}{3}\, {\rm R2sFL}[v]
+\frac{2}{3}\,{\rm R2sFH} [v]\right\} \exp\left(-\frac{s}{T^2}\right)\, ,
\end{eqnarray}
where the ${\rm R2sFF} [v]$, ${\rm R2sFA} [v]$, ${\rm R2sFL} [v]$ and
${\rm R2sFH} [v]$ with the variable  $v=\left(1-\frac{m_c^2}{s} \right)/\left(1+\frac{m_c^2}{s} \right)$ are mathematical functions defined at the energy-scale of the pole mass $\mu=m_c$, here the $n_l$ counts the number of massless quarks \cite{QCDSR-3loop}.

  We can derive Eq.(13) with respect to $1/T^2$, then eliminate the decay constant $f_{D_s}$ to obtain the QCD sum rules for the mass  $m_{D_s}$.
 The QCD sum rules for the decay constants and masses of the pseudoscalar  mesons $D$, $B$ and $B_s$ can be obtained with simple replacements.

\section{Numerical results and discussions}
The  masses  of the  pseudoscalar  mesons  listed in the   Review of Particle Physics are $m_{D^\pm}=(1869.5 \pm 0.4)\,\rm{MeV}$,  $m_{D^0}=(1864.91\pm 0.17)\,\rm{MeV}$, $m_{D_s^\pm}=(1969.0 \pm 1.4)\,\rm{MeV}$, $m_{B^\pm}=(5279.25\pm0.26)\,\rm{MeV}$,  $m_{B^0}=(5279.55\pm0.26)\,\rm{MeV}$, $m_{B_s^0}=(5366.7 \pm 0.4)\,\rm{MeV}$
\cite{PDG}. In 2010, the BaBar collaboration observed four excited charmed
mesons $D(2550)$, $D(2600)$, $D(2750)$ and $D(2760)$ in the decay
modes  $D^0(2550)\to D^{*+}\pi^-$, $D^{0}(2600)\to
D^{*+}\pi^-,\,D^{+}\pi^-$,  $D^0(2750)\to D^{*+}\pi^-$,
$D^{0}(2760)\to D^{+}\pi^-$, $D^{+}(2600)\to D^{0}\pi^+$ and
$D^{+}(2760)\to D^{0}\pi^+$ respectively in the inclusive $e^+e^-
\rightarrow c\bar{c}$  interactions  at the SLAC PEP-II
asymmetric-energy collider \cite{Babar2010}. The doublet $(D(2550),D(2600))$ are tentatively identified as the $2S$ doublet
$(0^-,1^-)$ \cite{D2550}.

We can take the threshold parameters  as $s^0_{D}=6.2\,\rm{GeV}^2$ and $s^0_{D_s}=7.3\,\rm{GeV}^2$ tentatively
 to avoid the contaminations of the high resonances, here we have taken into account the width of the  $D(2550)$ and the $SU(3)$ symmetry breaking effects.
If additional uncertainties $\delta s_0=0.5\,\rm{GeV}^2$ are supposed, then $\sqrt{s^0_{D}}-m_{D}=(0.5-0.7)\,\rm{GeV}$ and $\sqrt{s^0_{D_s}}-m_{D_s}=(0.6-0.8)\,\rm{GeV}$, the contributions of the ground states are fully included. In Ref.\cite{QCDSRfD5}, S. Narison takes the threshold parameters as $s^0_D=(5.3-9.5)\,\rm{GeV}^2$ and $s^0_B=(33-45)\,\rm{GeV}^2$. In this article, we take threshold parameters as $s^0_{B}=(33.5\pm 1.0)\,\rm{GeV}^2$ and $s^0_{B_s}=(35.0\pm 1.0)\,\rm{GeV}^2$ for the bottom mesons, the energy gaps are  $\sqrt{s^0_{B}}-m_{B}=(0.4-0.6)\,\rm{GeV}$ and $\sqrt{s^0_{B_s}}-m_{B_s}=(0.5-0.6)\,\rm{GeV}$, the contributions of the ground states are also fully  included.

The contaminations of the high resonances are very small if there are some contaminations.  We expect that the couplings of
 the pseudoscalar currents  to the excited states are more weak than that to the ground states. For example, the decay constants of the pseudoscalar mesons $\pi(140)$ and $\pi(1800)$ have the hierarchy $f_{\pi(1300)}\ll f_{\pi(140)}$ from the Dyson-Schwinger equation \cite{CDRoberts}, the lattice QCD \cite{Latt-pion},  the QCD sum rules \cite{QCDSR-pion}, etc, or from the experimental data \cite{pion-exp}. In fact, we can also choose  smaller threshold parameters, as the ground states $D$, $D_s$, $B$ and $B_s$ are very narrow, and search for the optimal values to reproduce the experimental values of the masses (In the case of {\bf II}, see Table 1 and related  paragraphs.).

The vacuum condensates are taken to be the standard values
$\langle\bar{q}q \rangle=-(0.25\pm0.01\, \rm{GeV})^3$, $\langle\bar{s}s \rangle=(0.8\pm0.1)\langle\bar{q}q \rangle$, $\langle \bar{q}g_s \sigma G q\rangle=m_0^2\langle\bar{q}q \rangle$, $\langle \bar{s}g_s \sigma G s\rangle=m_0^2\langle\bar{s}s \rangle$, $m_0^2=(0.8\pm0.1)\,\rm{GeV}^2$  at the energy scale  $\mu=1\, \rm{GeV}$ \cite{ColangeloReview}. The quark condensate evolves with the   renormalization group equation, $\langle\bar{q}q \rangle(\mu^2)=\langle\bar{q}q \rangle(Q^2)\left[\frac{\alpha_{s}(Q)}{\alpha_{s}(\mu)}\right]^{\frac{4}{9}}$.
 The value of the gluon condensate $\langle \frac{\alpha_s
GG}{\pi}\rangle $ has been updated from time to time, and changes
greatly \cite{NarisonBook}, we use the recently updated value $\langle \frac{\alpha_s GG}{\pi}\rangle=(0.022 \pm
0.004)\,\rm{GeV}^4 $ \cite{gg-conden}, and take the three-gluon condensate as $\langle g_s^3 GGG\rangle=(8.8\pm5.5)\,{\rm{GeV}^2}\langle\alpha_s GG\rangle=(0.616\pm0.385)\,\rm{GeV}^6$ \cite{gg-conden}. The recently updated value comes from the (Borel and moments) QCD sum rules study of the charmonium states   by including perturbative  corrections up to order ${\mathcal{O}}(\alpha_s^3)$ and vacuum condensates up to dimension $D=8$ \cite{gg-conden}, and it is superior  to the old value based on the low-order approximation in the operator product expansion.

Now, we take a short digression to discuss the relation between the pole mass and the $\overline{MS}$ mass. In QCD, the perturbative quark propagator in the momentum space can be written as
\begin{eqnarray}
S(p)&=&\frac{i}{\!\not\!{p}-m^0-\Sigma(\!\not\!{p},m^0)}\, ,
\end{eqnarray}
where the $m^0$ is the bare mass and the $\Sigma(\!\not\!{p},m^0)$ is the self-energy comes from the one-particle irreducible Feynman diagrams. The renormalized mass $m_r$ is defined as $m^0=m_r+\delta m$. It is convenient to choose the $\overline{MS}$ renormalization scheme by using the counterterm $\delta m$ to absorb the ultraviolet divergences of the form $\left[1/\epsilon+\log4\pi-\gamma_E\right]^L$, $L=1,2,\cdots$, then the $m_r$ is the $\overline{MS}$ mass. On the other hand, we can also define the pole mass by the setting $\!\not\!{p}-m^0-\Sigma(\!\not\!{p},m^0)=0$ with the on-shell mass $\!\not\!{p}=m$. The pole mass and the $\overline{MS}$ mass have the relation $m-m_r=\delta m+\Sigma(m,m^0)$. In QED, the electron mass is a directly observable quantity,  the pole mass is the physical mass and it is more convenient to choose the pole mass. While in QCD, the quark mass is not a directly observable quantity,  we have two choices in perturbative calculations.

In this article, we study the decay constants of the heavy pseudoscalar mesons with the following two possible choices:
\\
{\bf I } We choose the $\overline{MS}$ masses by setting $m=m(\mu)$ and take perturbative corrections up to the order $\mathcal{O}(\alpha_s)$. In other words, we take  the $R\left(\frac{m_Q^2}{s}\right)$ only;
\\
{\bf II} We choose the pole masses $m$, take perturbative corrections up to the order $\mathcal{O}(\alpha_s^2)$ and set the energy-scale $\mu=m_Q$.

The analytical expression of the  perturbative  $\mathcal{O}(\alpha_s)$ corrections $R\left(\frac{m_Q^2}{s}\right)$ is well known \cite{Aliev-pert}, while the
semi-analytical perturbative  $\mathcal{O}(\alpha_s^2)$ corrections are presented as mathematical functions ${\rm R2sFF} [v]$, ${\rm R2sFA} [v]$, ${\rm R2sFL} [v]$ and
$ {\rm R2sFH}[v]$ with the variable $v=\left(1-\frac{m_Q^2}{s} \right)/\left(1+\frac{m_Q^2}{s} \right)$ at the energy-scale of the heavy quark pole mass $\mu=m_Q$ \cite{QCDSR-3loop}. The analytical expressions of the  terms which contain  logarithms  such as $\log\frac{\mu^2}{m_Q^2}$, $\log\frac{\mu^2}{s}$ cannot be recovered,
   it is unreasonable to take other energy scale besides $m_Q$. We have to set $\mu=m_Q$, if the semi-analytical perturbative  $\mathcal{O}(\alpha_s^2)$ corrections are taken into account.

In the case of {\bf I}, we take the $\overline{MS}$ masses $m_{c}(m_c^2)=(1.275\pm0.025)\,\rm{GeV}$,  $m_{b}(m_b^2)=(4.18\pm 0.03)\,\rm{GeV}$,  $m_s(\mu=2\,\rm{GeV})=(0.095\pm0.005)\,\rm{GeV}$
 from the Particle Data Group \cite{PDG}, and set $m_q=0$. Furthermore, we take into account
the energy-scale dependence of  the $\overline{MS}$ masses from the renormalization group equation,
\begin{eqnarray}
m_s(\mu^2)&=&m_s({\rm 4GeV}^2 )\left[\frac{\alpha_{s}(\mu)}{\alpha_{s}({\rm 2GeV})}\right]^{\frac{4}{9}} \, ,\nonumber\\
m_c(\mu^2)&=&m_c(m_c^2)\left[\frac{\alpha_{s}(\mu)}{\alpha_{s}(m_c)}\right]^{\frac{12}{25}} \, ,\nonumber\\
m_b(\mu^2)&=&m_b(m_b^2)\left[\frac{\alpha_{s}(\mu)}{\alpha_{s}(m_b)}\right]^{\frac{12}{23}} \, ,\nonumber\\
\alpha_s(\mu)&=&\frac{1}{b_0t}\left[1-\frac{b_1}{b_0^2}\frac{\log t}{t} +\frac{b_1^2(\log^2{t}-\log{t}-1)+b_0b_2}{b_0^4t^2}\right]\, ,
\end{eqnarray}
  where $t=\log \frac{\mu^2}{\Lambda^2}$, $b_0=\frac{33-2n_f}{12\pi}$, $b_1=\frac{153-19n_f}{24\pi^2}$, $b_2=\frac{2857-\frac{5033}{9}n_f+\frac{325}{27}n_f^2}{128\pi^3}$,  $\Lambda=213\,\rm{MeV}$, $296\,\rm{MeV}$  and  $339\,\rm{MeV}$ for the flavors  $n_f=5$, $4$ and $3$, respectively  \cite{PDG}. For the $D$ ($D_s$) mesons, we take  $n_f=3$ and $\mu=\sqrt{m_D^2-m_c^2}\approx 1\,\rm{GeV}$; for the $B$ ($B_s$) mesons, we take $n_f=4$ and $\mu=\sqrt{m_B^2-m_b^2}\approx 2.5\,\rm{GeV}$.

In the case of {\bf II}, we take the pole masses,  set $n_f=4$ and $\mu=m_c$ for the $D$ ($D_s$) mesons and $n_f=5$ and $\mu=m_b$ for the $B$ ($B_s$) mesons.

Firstly, we study the masses and decay constants of the heavy pseudoscalar mesons  in the case of {\bf I}.

In Fig.6, we plot the contributions of different terms in the operator product expansion with variations of the Borel parameters. From the figure, we can see that the convergence of the operator product expansion cannot be satisfied for the $D$ and $D_s$ ($B$ and $B_s$) mesons  at the region $T^2< 0.9\,\rm{GeV}^2$ ($T^2< 3.0\,\rm{GeV}^2$). In Figs.7-8,
we plot the masses and decay constants with variations of the Borel parameters at large ranges. Although there
appear minimum platforms for the masses and decay constants of the $D$ and $D_s$ mesons at $T^2\leq 0.9\,\rm{GeV}^2$, the Borel windows cannot be chosen in such regions.
For the $B$ and $B_s$ mesons, the decay constants decrease monotonously with increase of the Borel parameter at $T^2<4\,\rm{GeV}^2$. We choose the suitable Borel parameters to satisfy the two criteria (pole dominance and convergence of the operator product expansion) of the QCD sum rules, and reproduce the experimental values
of the masses. The vacuum condensates play a less important role in the Borel windows, but they  play an important role in determining the Borel windows.
The threshold parameters, Borel parameters, pole contributions and the resulting decay constants are shown explicitly in Table 1.

In calculations, we observe that the ground state masses are sensitive to the heavy quark $\overline{MS}$ masses, i.e. they
 increase  monotonously with increase  of the heavy quark $\overline{MS}$ masses. The $\overline{MS}$
 masses from the Particle Data Group  happen to result in  satisfactory ground state masses compared to the experimental data \cite{PDG}.

In Table 2, we compare the present predictions  to  the experimental data and some (not all) theoretical calculations. The value $f_B=(194\pm9)\,\rm{MeV}$ listed in the Review of Particle Physics \cite{PDG} is  the average of the lattice QCD calculations \cite{LattfD3,fB-exp-PDG}. The present predictions $f_D=(208\pm11)\,\rm{MeV}$ and $f_B=(189\pm15)\,\rm{MeV}$ are consistent with the experimental data within uncertainties, while the prediction  $f_{D_s}=(241\pm12)\,\rm{MeV}$ is below the lower bound of the experimental data $f_{D_s}=(260.0\pm5.4)\,\rm{MeV}$ \cite{PDG}. The ratio $f_{D_s}/f_D\approx f_{B_s}/f_B$, the heavy quark symmetry works well.
In the early work \cite{Khlopov-2},   Gershtein  and
 Khlopov obtained a simple relation $f_{ij}\propto m_i+m_j$ for the decay constant $f_{ij}$ of the pseudoscalar meson having the constituent quarks $i$ and $j$,
 the simple relation does not work well enough numerically.

Secondly, we study the masses and decay constants of the heavy pseudoscalar mesons  in the case of {\bf II}.

The values of the pole masses listed in the Review of Particle Physics are $m_c=(1.67\pm 0.07)\,\rm{GeV}$ and $m_b=(4.78\pm0.06)\,\rm{GeV}$  \cite{PDG}, which correspond to the $\overline{MS}$ masses $m_{c}(m_c^2)=(1.275\pm0.025)\,\rm{GeV}$ and $m_{b}(m_b^2)=(4.18\pm 0.03)\,\rm{GeV}$, respectively. In calculations, we observe that
the heavy pseudoscalar meson masses increase  monotonously with increase  of the pole masses, the  values  $m_c=1.67\,\rm{GeV}$ and $m_b=4.78\,\rm{GeV}$ cannot lead to satisfactory results by choosing reasonable Borel parameters and threshold parameters. We expect that smaller pole masses maybe lead to
satisfactory heavy pseudoscalar meson masses, and search for the optimal values.

 In Fig.9, we plot the predicted  masses  with variations of the pole masses with the threshold parameters $s^0_D=5.5\,\rm{GeV}^2$, $s^0_{D_s}=7.2\,\rm{GeV}^2$, $s^0_B=32.0\,\rm{GeV}^2$, $s^0_{B_s}=34.5\,\rm{GeV}^2$ and  Borel parameters $T^2_{D}=1.7\,\rm{GeV}^2$, $T^2_{D_s}=1.3\,\rm{GeV}^2$, $T^2_{B}=4.5\,\rm{GeV}^2$, $T^2_{B_s}=4.7\,\rm{GeV}^2$ at large ranges.
In Fig.10, we plot the corresponding  decay constants  with variations of the pole masses with the same parameters as in Fig.9.
From Fig.9, we can see that the pole masses $m_c=1.47\,\rm{GeV}$ and $m_b=4.64\,\rm{GeV}$ are the optimal values to reproduce the experimental values of the heavy meson masses.   Detailed analysis  indicates that those threshold parameters and Borel parameters are also optimal values.

In this article, we  take the pole masses as $m_c=(1.47\pm 0.06)\,\rm{GeV}$ and $m_b=(4.64\pm0.06)\,\rm{GeV}$, which lead to the uncertainties $\delta m_D=\pm 0.03\,\rm{GeV}$, $\delta m_{D_s}=\pm 0.05\,\rm{GeV}$, $\delta m_B=\pm 0.04\,\rm{GeV}$ and $\delta m_{B_s}=\pm 0.04\,\rm{GeV}$. The uncertainties of $\pm(0.03-0.05)\,\rm{GeV}$ are acceptable in the QCD sum rules.  In Ref.\cite{SRfDEFT1}, Penin and  Steinhauser take the $b$-quark pole mass as $m_b=4.68\,\rm{GeV}$ ($4.78\,\rm{GeV}$), and study the decay constant $f_B$ by including perturbative ${\mathcal{O}}(\alpha_s)$ (${\mathcal{O}}(\alpha_s^2)$) corrections with the QCD sum rules in the heavy quark effective theory, then estimate the $c$-quark pole mass to be $m_c=(1.37\pm 0.10)\,\rm{GeV}$. The  values of the pole masses $m_Q$ from different references overlap with (but not equal to) each other.
One may expect to choose larger uncertainties of the pole masses, however, larger $\delta m_c$ and $\delta m_b$ mean larger derivations from the experimental data $m_{D^\pm}=1869.5 \,\rm{MeV}$,  $m_{D^0}=1864.91\,\rm{MeV}$, $m_{D_s^\pm}=1969.0 \,\rm{MeV}$, $m_{B^\pm}=5279.25\,\rm{MeV}$,  $m_{B^0}=5279.55\,\rm{MeV}$ and  $m_{B_s^0}=5366.7 \,\rm{MeV}$ \cite{PDG}, see Fig.9. The light quark masses play less important roles, we take the values $m_u=m_d=0$ and $m_s=0.15\,\rm{GeV}$.

Once  the pole masses are fixed, we choose   suitable Borel parameters and threshold parameters to satisfy the two criteria (pole dominance and convergence of the operator product expansion) of the QCD sum rules, and reproduce the experimental values
of the masses. The threshold parameters, Borel parameters, pole contributions and the resulting decay constants are also shown explicitly in Table 1. The resulting decay constants  are compared to  the experimental data and some (not all) theoretical calculations in Table 2.

From Table 2, we can see that the present  predictions $f_D=(211\pm14)\,\rm{MeV}$ and $f_B=(190\pm17)\,\rm{MeV}$ are consistent with the experimental data within uncertainties, while the prediction  $f_{D_s}=(258\pm13)\,\rm{MeV}$ is in excellent agreement with  the experimental data $f_{D_s}=(260.0\pm5.4)\,\rm{MeV}$ within uncertainties \cite{PDG}. Furthermore, $f_{D_s}/f_D\approx f_{B_s}/f_B$, the heavy quark symmetry works well, the ratio $f_{D_s}/f_D=1.22\pm0.08$ is in excellent agreement with  the experimental data $f_{D_s}/f_D=1.26\pm0.06$ \cite{PDG}; while most of the theoretical predictions (including the present prediction in the case of {\bf I}) of the ratio $f_{D_s}/f_D$ are below the experimental data.  We can draw the conclusion tentatively that the prediction $f_{B_s}=(233\pm17)\,\rm{MeV}$ is robust.

In Ref.\cite{QCDSRfD3},  A. Khodjamirian estimates the upper bound $f_D<230 \,\rm{MeV}$ and $f_{D_s}<270 \,\rm{MeV}$ based on the QCD sum rules, the present predictions (both in the cases  of {\bf I} and {\bf II}) satisfy the constraints. The differences between the predictions  in the cases of  {\bf I} and {\bf II} originate from the systematic uncertainties of the QCD sum rules, we cannot come to the conclusion which predictions are the true values or more close to the true values.
The existence of a charged Higgs boson or any other charged
object beyond the standard model would modify the decay rates,  therefore   modify the values of the decay constants, for example, the leptonic decay widths are modified in two-Higgs-doublet models \cite{Higgs}. If the predictions in the case of {\bf I} are more close to the true values, new physics beyond the standard model are favored so as to smear the discrepancies  between the theoretical calculations and experimental data. On the other hand, if the predictions in the case of {\bf II} are more close to the true values, new physics beyond the standard model are not favored, as the agreements between the experimental data and present theoretical calculations are already excellent.

In the QCD sum rules, the resulting ground state masses are sensitive to the heavy quark masses, variations of the heavy quark masses lead to changes of integral ranges $(m_Q+m_q)^2-s_0$ of the variable  $ \rm \bf {ds}$ besides the QCD spectral densities, therefore changes of the Borel windows and predicted masses and decay constants. In this article, we choose both the $\overline{MS}$ masses and pole masses to study the masses and decay constants of the heavy pseudoscalar mesons.
We take the following criteria:

$\bullet$ Pole dominance at the phenomenological side;

$\bullet$ Convergence of the operator product expansion;

$\bullet$ Appearance of the Borel platforms;

$\bullet$ Reappearance  of experimental values of the ground state heavy meson masses.

The values of the heavy quark  $\overline{MS}$ masses from the Particle Data Group can satisfy the four  criteria; while the values of the heavy quark  pole masses
from the Particle Data Group cannot  satisfy the four  criteria, we choose smaller but reasonable pole masses to satisfy the four  criteria. The $\overline{MS}$ masses and pole masses lead to quite different decay constants for the $D_s$ and $B_s$ mesons. Recently, the Belle collaboration extracted the value,
\begin{eqnarray}
 f_{D_s}&=& (255.5 \pm 4.2 \pm 5.1)\,\rm{ MeV}\,,
\end{eqnarray}
from the decays $D_s^+ \to \mu^+ \nu_\mu $ and $D_s^+ \to \tau^+ \nu_\tau $ \cite{Belle1307}, which is in excellent agreement with the present prediction $f_{D_s}=(258\pm13)\,\rm{MeV}$ in the case $\bf \rm II$.

\begin{table}
\begin{center}
\begin{tabular}{|c|c|c|c|c|c|c|c|}\hline\hline
              & $T^2 (\rm{GeV}^2)$  & $s_0 (\rm{GeV}^2)$   & pole         & $m_P(\rm{GeV})$    & $f_{P}(\rm{MeV})$   \\ \hline
  $D$ (I)     & $1.2-1.8$           & $6.2\pm0.5$          & $(66-93)\%$  & $1.87\pm0.10$      & $208\pm11$          \\ \hline
 $D_{s}$ (I)  & $1.2-1.7$           & $7.3\pm0.5$          & $(80-96)\%$  & $1.97\pm0.10$      & $241\pm12$          \\ \hline
   $B$ (I)    & $4.7-6.1$           & $33.5\pm1.0$         & $(46-70)\%$  & $5.28\pm0.06$      & $189\pm15$          \\ \hline
 $B_{s}$ (I)  & $5.1-6.6$           & $35.0\pm1.0$         & $(47-71)\%$  & $5.37\pm0.06$      & $216\pm16$          \\ \hline
  $D $ (II)   & $1.4-2.0$           & $5.5\pm0.5$          & $(54-84)\%$  & $1.87\pm0.06$      & $211\pm14$          \\ \hline
 $D_{s}$ (II) & $1.0-1.6$           & $7.2\pm0.5$          & $(84-99)\%$  & $1.97\pm0.08$      & $258\pm13$          \\ \hline
   $B$ (II)   & $4.1-4.9$           & $32.0\pm1.0$         & $(51-71)\%$  & $5.28\pm0.06$      & $190\pm17$          \\ \hline
 $B_{s}$ (II) & $4.3-5.1$           & $34.5\pm1.0$         & $(61-79)\%$  & $5.37\pm0.05$      & $233\pm17$          \\ \hline
 \hline
\end{tabular}
\end{center}
\caption{ The Borel parameters, continuum threshold parameters, pole contributions, masses and decay constants for the heavy pseudoscalar mesons. }
\end{table}

\begin{figure}
 \centering
 \includegraphics[totalheight=5cm,width=6cm]{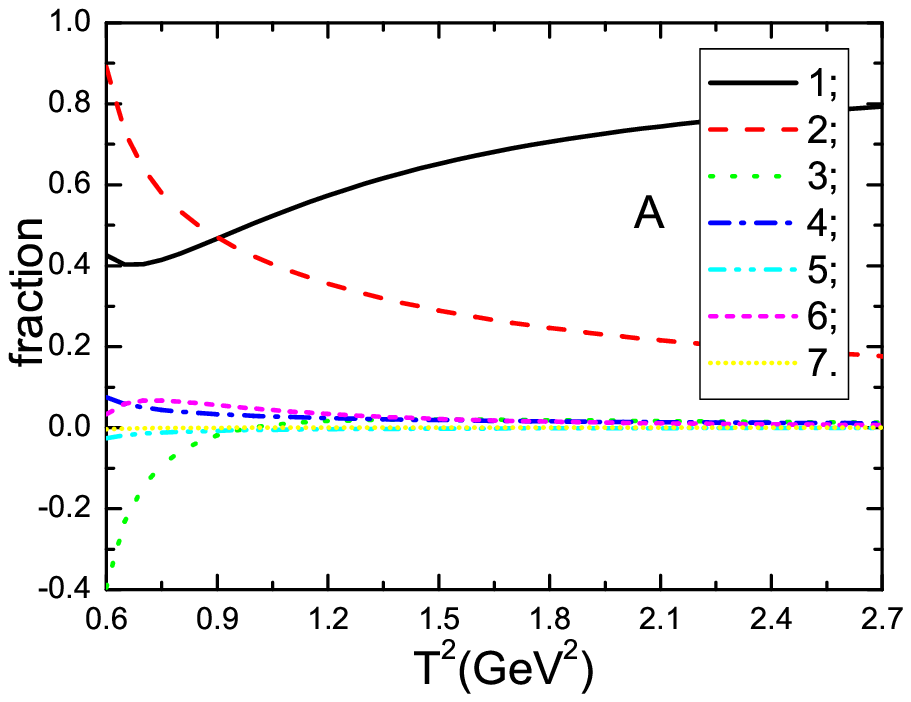}
\includegraphics[totalheight=5cm,width=6cm]{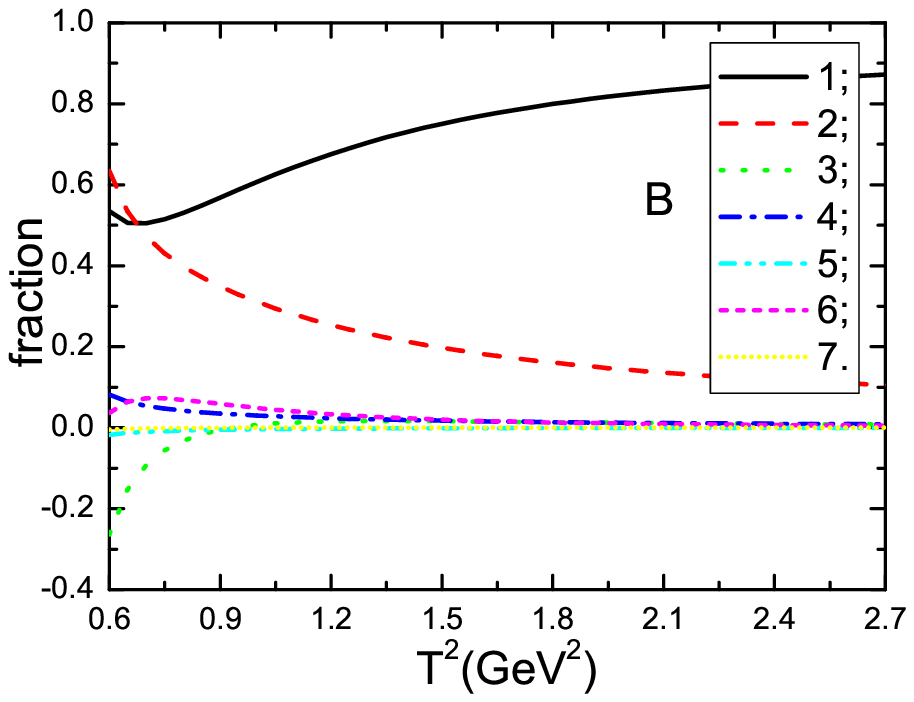}
\includegraphics[totalheight=5cm,width=6cm]{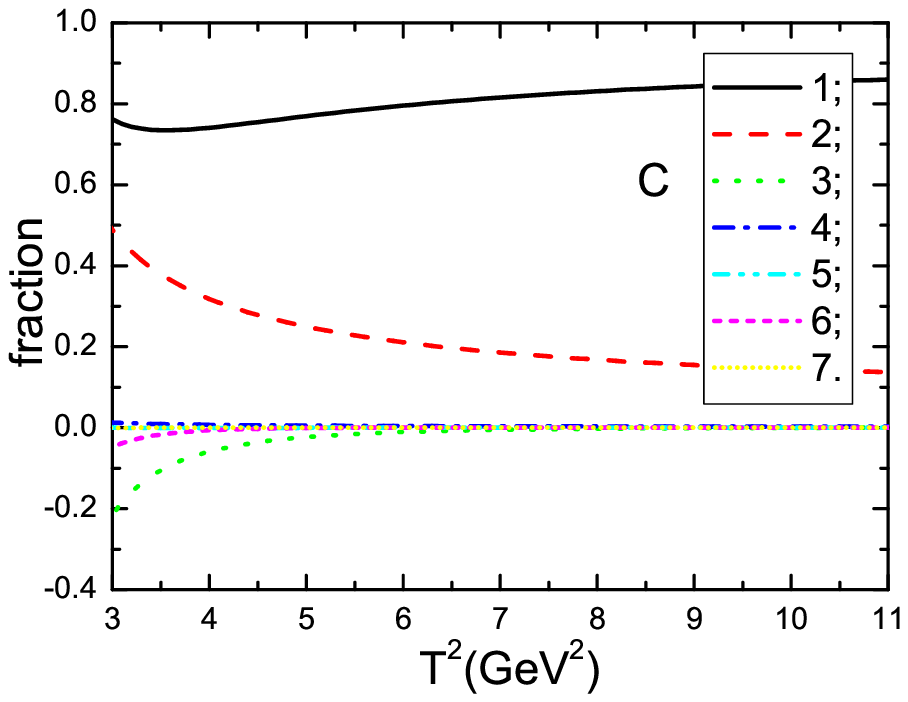}
\includegraphics[totalheight=5cm,width=6cm]{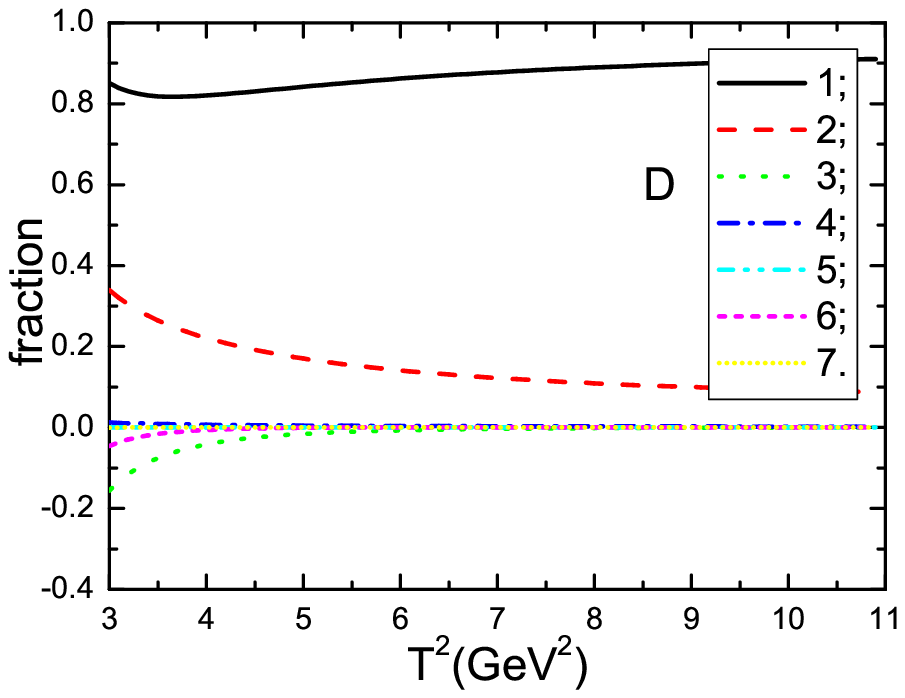}
        \caption{ The contributions of different terms in the operator product expansion, the $A$, $B$, $C$ and $D$ correspond to the $D$, $D_s$, $B$ and $B_s$, respectively; while the 1, 2, 3, 4, 5, 6 and 7 denote the contributions of the perturbative terms, quark condensate, mixed quark condensate, gluon condensate,
        four-quark condensate $\mathcal{O}(\alpha_s)$, three-gluon condensate and four-quark condensate $ \mathcal{O}(\alpha_s^2)$, respectively.}
\end{figure}

\begin{table}
\begin{center}
\begin{tabular}{|c|c|c|c|c|c|c|}\hline\hline
                      &$f_{D}(\rm{MeV})$     &$f_{D_{s}}(\rm{MeV})$ &$f_{B}(\rm{MeV})$     &$f_{B_{s}}(\rm{MeV})$ &$f_{D_{s}}/f_{D}$     &$f_{B_{s}}/f_{B}$\\ \hline
 Expt \cite{PDG}      &$206.7\pm8.9$         &$260.0\pm5.4$         &$194\pm9$             &                      &$1.26\pm0.06$         &                \\ \hline
 QCDSR \cite{SRfDEFT1}&$195\pm 20$           &                      &$206 \pm 20 $         &                      &                      &               \\ \hline
 QCDSR \cite{QCDSRfD2}&$177\pm 21$           &$205\pm22$            &$178 \pm 14 $         &$200\pm14$            &$1.16\pm0.16$         &$1.12\pm0.11$        \\ \hline
 QCDSR \cite{QCDSRfD4}&$206.2\pm7.3$         &$245.3\pm15.7$        &$193.4\pm12.3$        &$232.5\pm18.6$        &$1.193\pm0.025$       &$1.203\pm0.020$   \\ \hline
 QCDSR \cite{QCDSRfD5}&$204\pm6$             &$246\pm6$             &$207\pm8$             &$234\pm5$             &$1.21\pm0.04$         &$1.14\pm0.03$   \\ \hline
 LQCD \cite{LattfD1}  &$197\pm9$             &$244\pm8$             &                      &                      &$1.24\pm0.03  $       &                 \\ \hline
 LQCD \cite{LattfD2}  &$213\pm4$             &$248.0\pm2.5$         &$191\pm9$             &$228\pm10$            &$1.164\pm0.018$       &$1.188\pm0.018$ \\ \hline
 LQCD \cite{LattfD3}  &$218.9\pm11.3$        &$260.1\pm10.8$        &$196.9\pm8.9$         &$242.0\pm9.5$         &$1.188\pm0.025$       &$1.229\pm0.026$ \\ \hline
 BSE \cite{BSEfD1}    &$238$                 &$241$                 &$193$                 &$195$                 &$1.01$                &$1.01$ \\ \hline
 RPM \cite{RPMfD2}    &$234$                 &$268$                 &$189$                 &$218$                 &$1.15$                &$1.15$ \\ \hline
 FCM \cite{FCMfD1}    &$210\pm10$            &$260\pm10$            &$182\pm8$             &$216\pm8$             &$1.24\pm0.03$         &$1.19\pm0.02$ \\ \hline
 LFQM\cite{LFQMfD2}   &$205.8\pm8.9$         &$264.5\pm17.5$        &$204\pm31$            &$270.0\pm42.8$        &$1.29\pm0.07$         &$1.32\pm0.08$ \\ \hline
 This work (I)        &$208\pm11$            &$241\pm12$            &$189\pm15$            &$216\pm16$            &$1.16\pm0.07$         &$1.14\pm0.11$   \\ \hline
 This work (II)       &$211\pm14$            &$258\pm13$            &$190\pm17$            &$233\pm17$            &$1.22\pm0.08$         &$1.23\pm0.12$   \\ \hline
 \hline
\end{tabular}
\end{center}
\caption{ The decay constants of the heavy pseudoscalar mesons from the  experimental data and  some theoretical calculations.}
\end{table}

\begin{figure}
 \centering
 \includegraphics[totalheight=5cm,width=6cm]{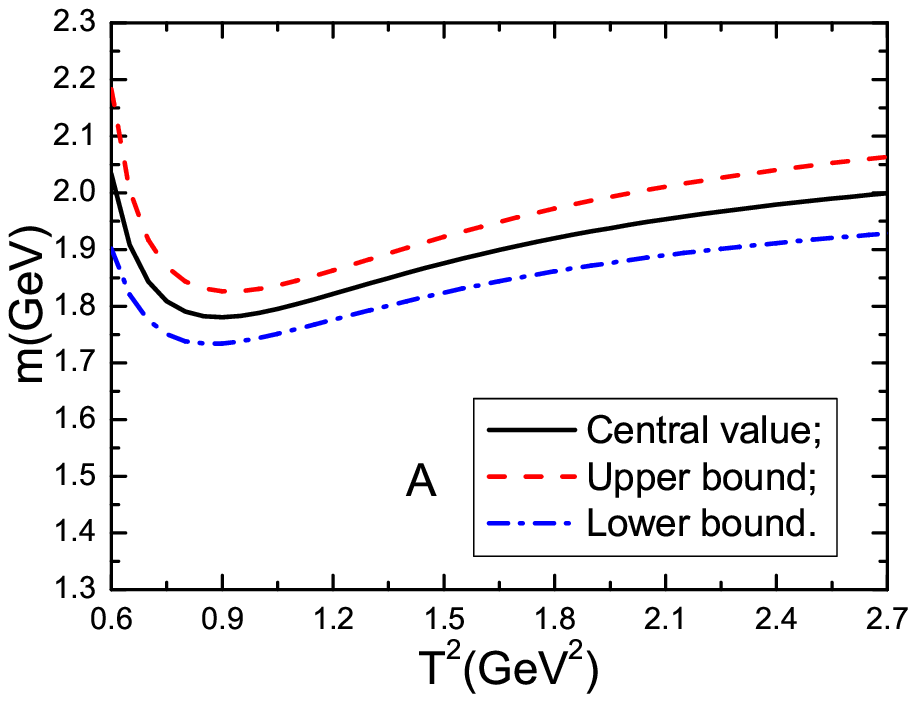}
\includegraphics[totalheight=5cm,width=6cm]{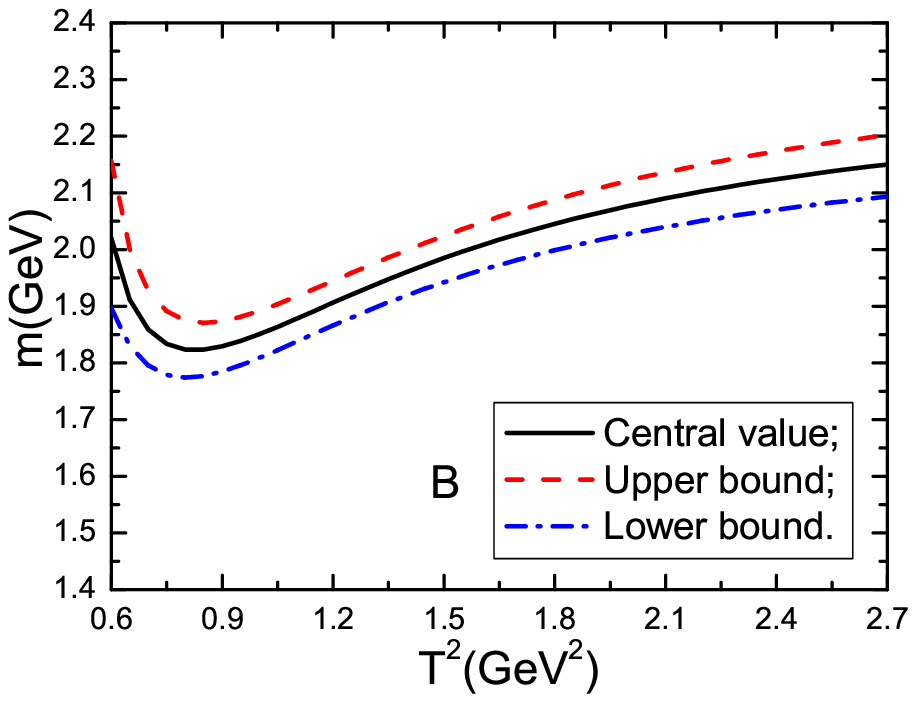}
\includegraphics[totalheight=5cm,width=6cm]{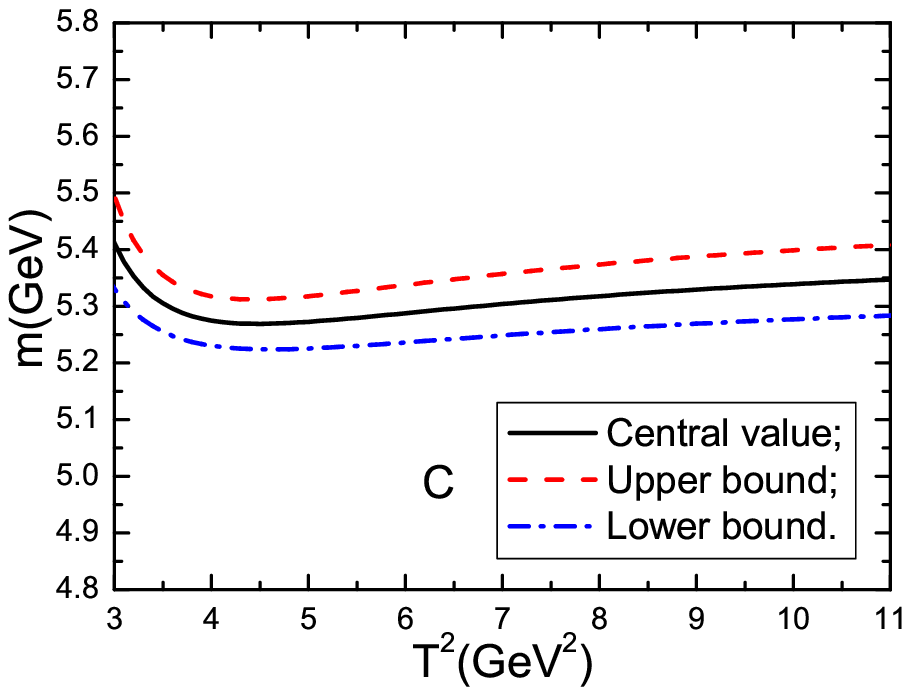}
\includegraphics[totalheight=5cm,width=6cm]{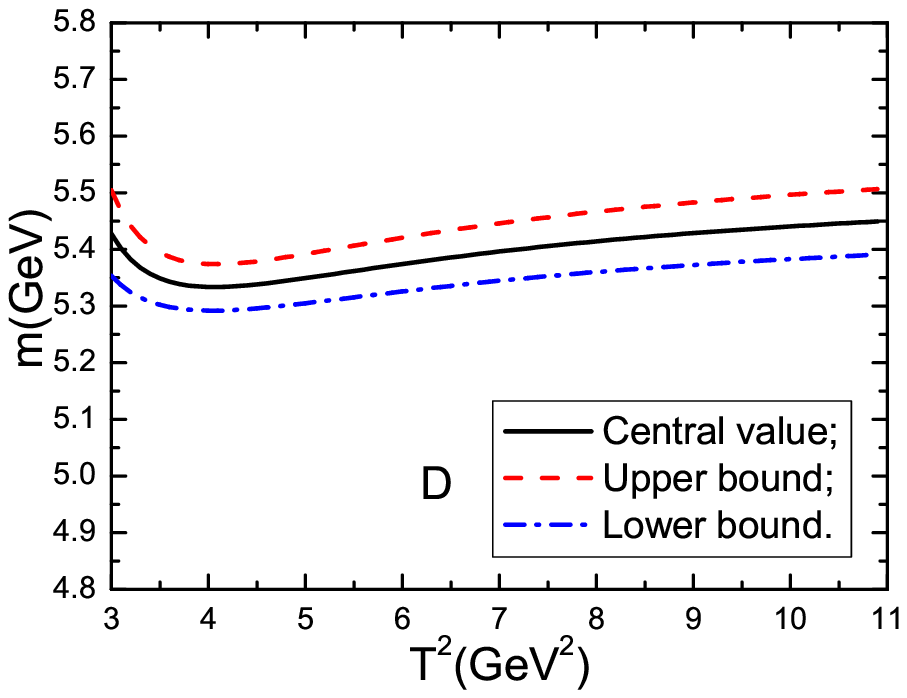}
        \caption{ The masses of the heavy pseudoscalar mesons, the $A$, $B$, $C$ and $D$ correspond to the $D$, $D_s$, $B$ and $B_s$, respectively.}
\end{figure}

\begin{figure}
 \centering
 \includegraphics[totalheight=5cm,width=6cm]{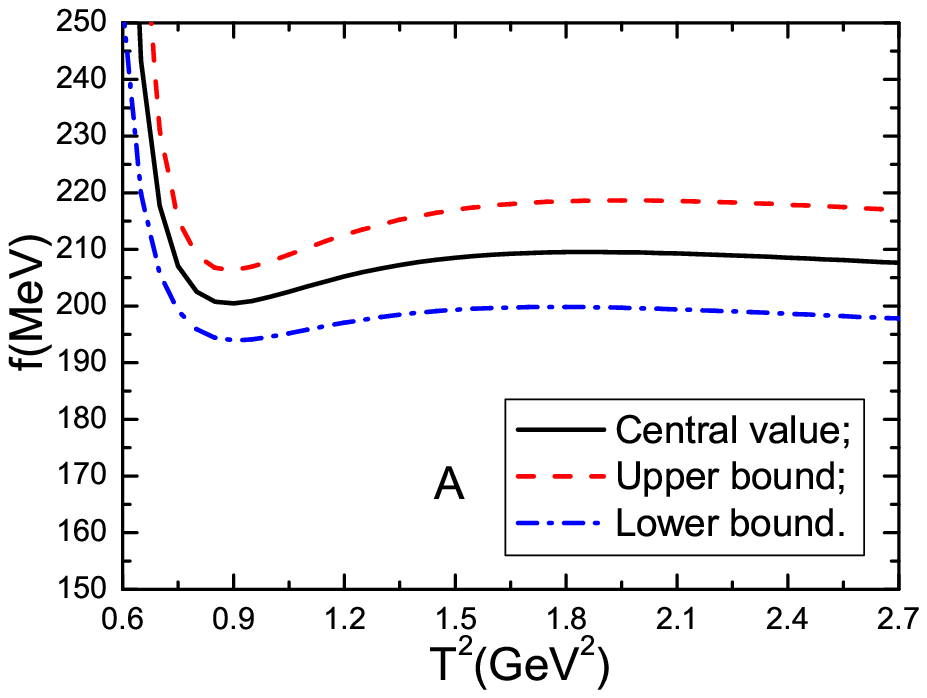}
\includegraphics[totalheight=5cm,width=6cm]{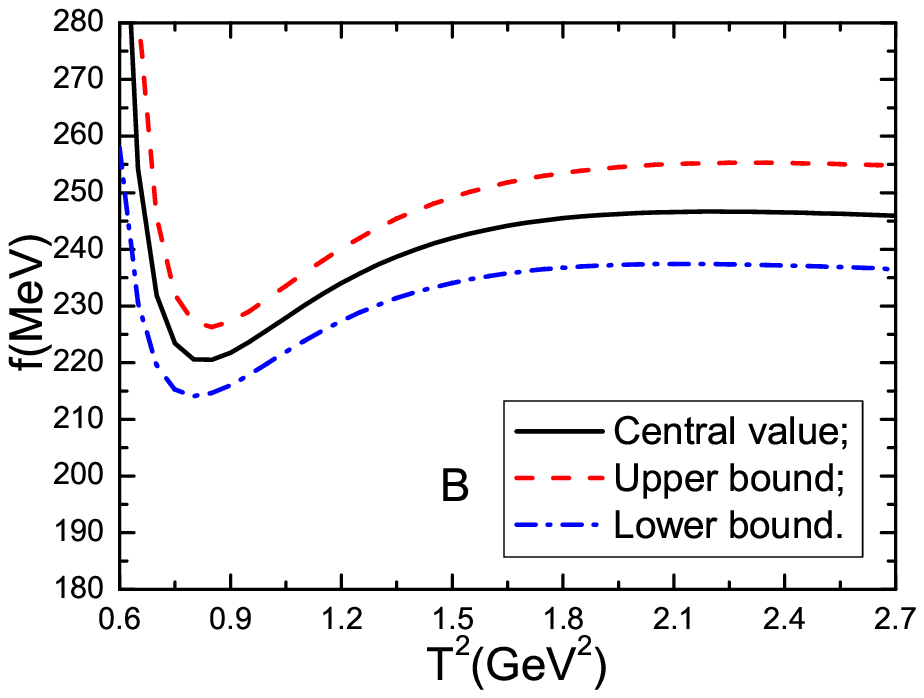}
\includegraphics[totalheight=5cm,width=6cm]{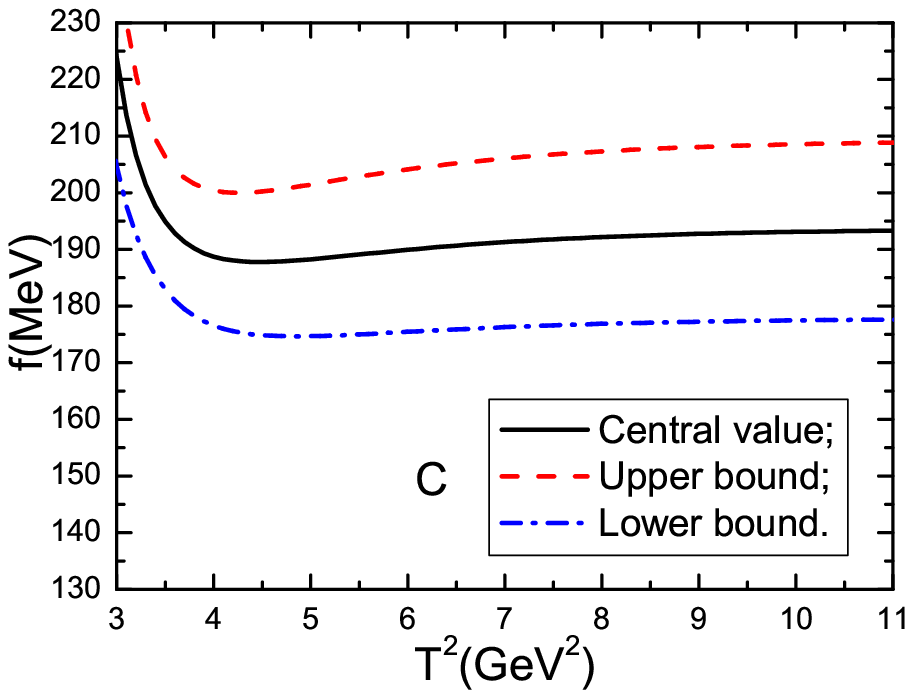}
\includegraphics[totalheight=5cm,width=6cm]{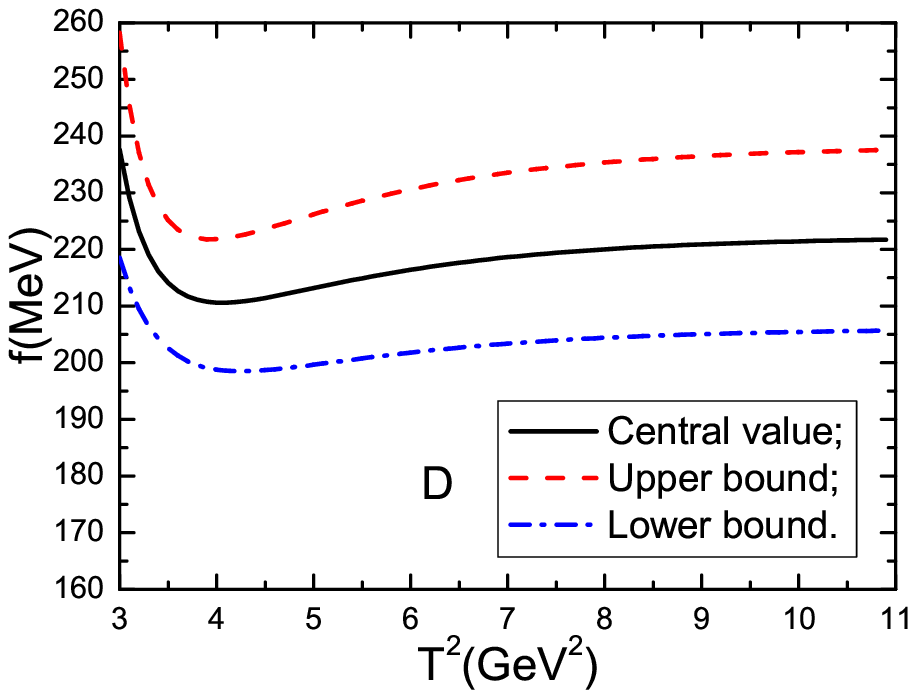}
        \caption{ The decay constants of the heavy pseudoscalar mesons, the $A$, $B$, $C$ and $D$ correspond to the $D$, $D_s$, $B$ and $B_s$, respectively.}
\end{figure}

\begin{figure}
 \centering
 \includegraphics[totalheight=5cm,width=6cm]{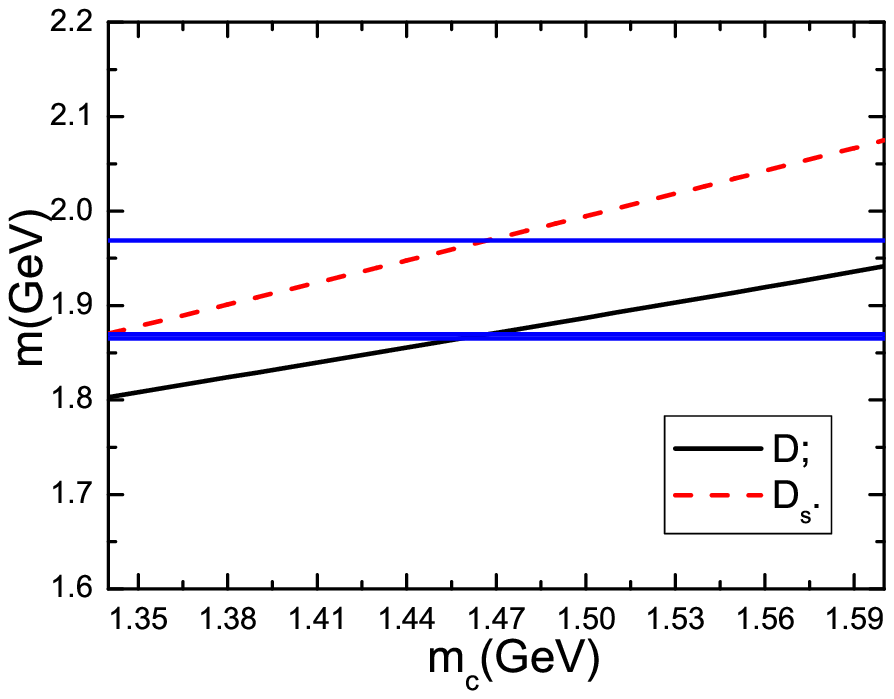}
\includegraphics[totalheight=5cm,width=6cm]{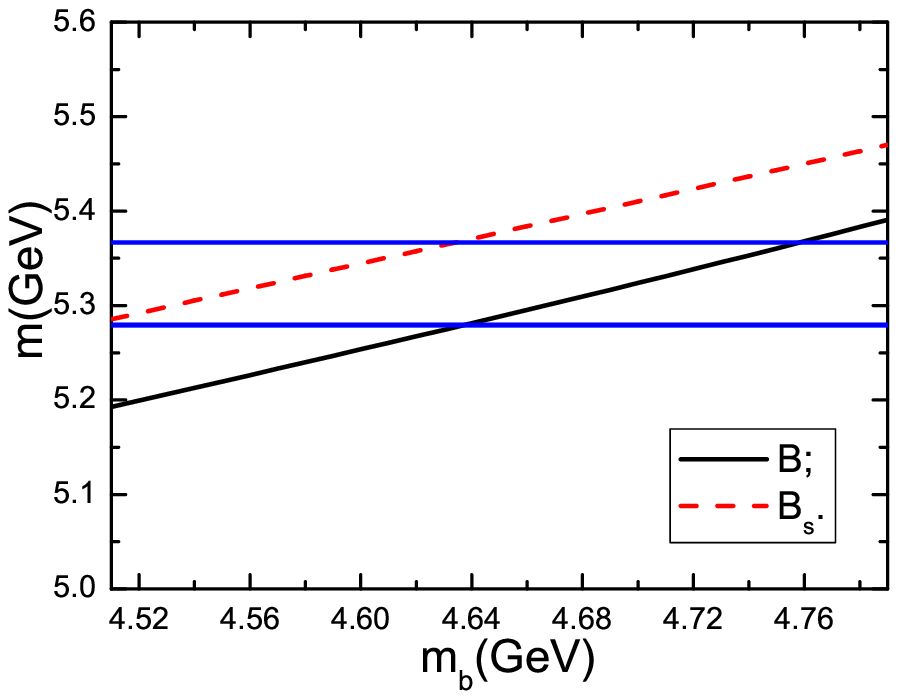}
        \caption{ The masses of the heavy pseudoscalar mesons with variations of the pole masses $m_c$ and $m_b$, where the horizontal lines denote  the experimental values.}
\end{figure}

\begin{figure}
 \centering
 \includegraphics[totalheight=5cm,width=6cm]{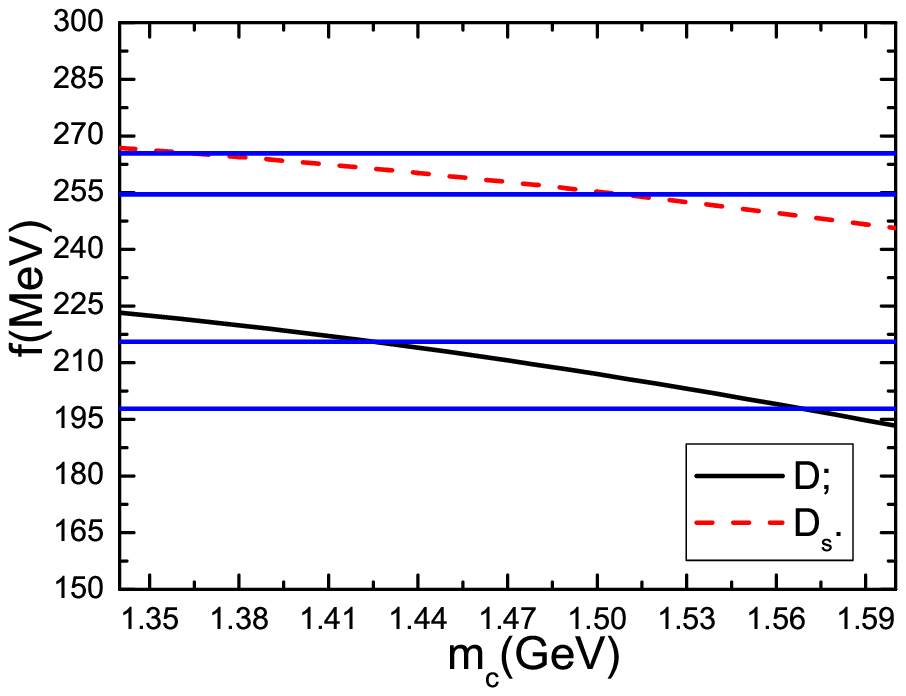}
\includegraphics[totalheight=5cm,width=6cm]{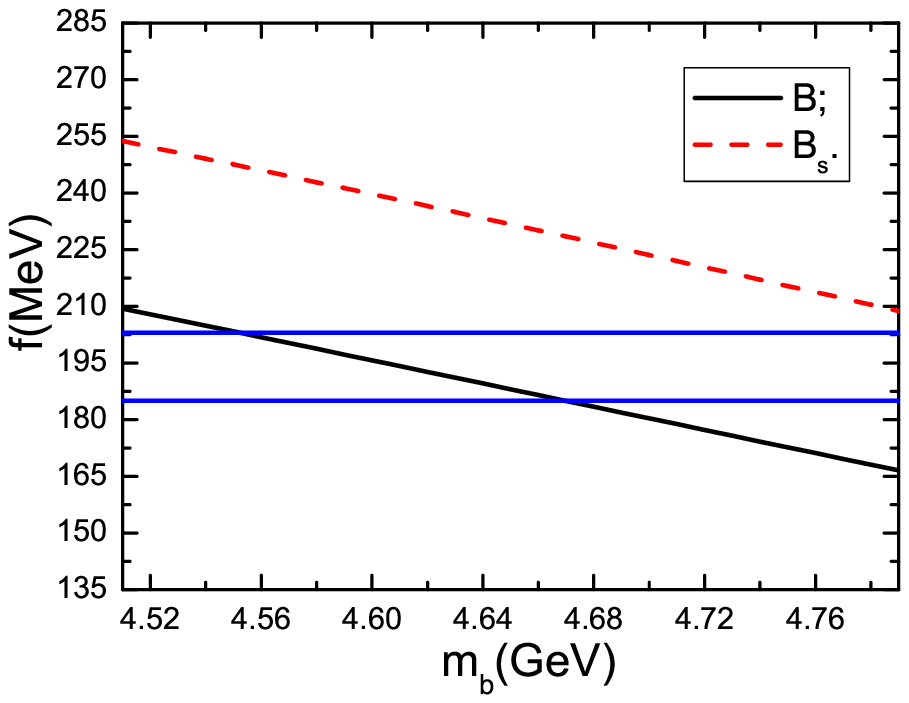}
        \caption{ The decay constants of the heavy pseudoscalar mesons with variations of the pole masses $m_c$ and $m_b$, where the horizontal lines denote the upper and lower bounds of the experimental values.}
\end{figure}

\section{Conclusion}
In this article, we recalculate the contributions of all vacuum condensates up to dimension-6, in particular the one-loop corrections  to the quark condensates $\alpha_s\langle \bar{q}q\rangle$ and partial one-loop corrections to the four-quark condensates  $\alpha_s^2\langle \bar{q}q \rangle^2$, in the operator product expansion, and obtain the analytical QCD spectral densities. Then we study the masses and decay constants of the heavy pseudoscalar mesons using the QCD sum rules
 with the two possible choices:
{\bf I } we choose the $\overline{MS}$ masses by setting $m=m(\mu)$ and take perturbative corrections up to the order $\mathcal{O}(\alpha_s)$;
{\bf II} we choose the pole masses $m$, take perturbative corrections up to the order $\mathcal{O}(\alpha_s^2)$ and set the energy-scale $\mu=m_Q$.
In the case of {\bf I}, the predictions  $f_D=(208\pm11)\,\rm{MeV}$ and $f_B=(189\pm15)\,\rm{MeV}$ are consistent with the experimental data within uncertainties, while the prediction $f_{D_s}=(241\pm12)\,\rm{MeV}$ is still below the lower bound of the experimental data $f_{D_s}=(260.0\pm5.4)\,\rm{MeV}$, new physics beyond the standard model  are favored so as to smear the discrepancies, in other words,  there are rooms for new physics to smear the discrepancies.
In the case of {\bf II}, the predictions $f_D=(211\pm14)\,\rm{MeV}$, $f_B=(190\pm17)\,\rm{MeV}$, $f_{D_s}=(258\pm13)\,\rm{MeV}$ and $f_{D_s}/f_D=1.22\pm0.08$ are all in excellent agreements  with the experimental data  within uncertainties,  new physics beyond the standard model are not favored, in other words, the new physics models should satisfy more stringent constraints. The differences between the predictions  in the cases of {\bf I} and {\bf II} originate from the systematic uncertainties of the QCD sum rules, we cannot come to the conclusion which predictions are the true values or more close to the true values.

\section*{Acknowledgements}
This  work is supported by National Natural Science Foundation,
Grant Number 11075053,  and the Fundamental Research Funds for the
Central Universities.

\end{document}